# Ten Simple Rules When Considering Retirement

Philip E. Bourne, PhD

You know a field is maturing when its early proponents think about retiring, or alas, pass away. So it is with computational biology. At 65, I think about retiring more. Not so much about retirement per se, but as in terms of what I want to accomplish before I retire and what does retirement mean to me in the first place. In other words, retirement is complex and these ten rules probably (and hopefully) just start a discourse. For most scientists, including computational biologists, it's not a situation of, now that you're 65, please accept your plaque, exit stage left, and goodbye. The questions I'm now considering are far more varied and nuanced. It's about where I want to focus my life and my energies. How much do I want to continue to mentor younger folk? How do I keep ties with colleagues I like and respect? How can I give back as much as possible to society and a profession that has treated me so well?

The focus here is on retirement as an option, not a requirement. After all emeritus status in academia, government and industry can typically go on in some form indefinitely.

It must also be said, ahead of the rules themselves, that in drafting the rules and having them reviewed by those acknowledged below, as well as discussing them with colleagues, friends and family, that I was entering a very personal space. Efforts to coopt co-authors, which I like to do to provide a broader perspective, resulted in something too diffuse. How individuals think, or indeed choose not to think, about retirement is very personal. As such the rules are more personal than I would normally write, or have written, on other subjects. Therefore, it may be that these rules do not resonate with you directly, but I hope they will at least make you think about what retirement means to you.

Rule 1: Consider what you want to do when you get up in the morning

This would seem key and well covered in the endless retirement literature. This defines your new days. Currently, I rise at around 6am, make coffee and then start on email, preparing a talk or class, or editing a paper or grant for a couple of hours before starting my official day. I have done that for many years. What would I fill my days with to replace that? Would I keep doing some of it? Would I sleep in, which I have not done for years? Would I read, go out walking, go motorcycling, activities I do now, but want to do more of? Would I start new activities?

My current thought is to start new activities before any major life change with enough vigor to determine whether I like them or not. It seems to me that starting a new lifestyle overnight could be a mistake; such drastic change would be jarring. Instead, it seems sensible to experiment with different options, figure out what you like and spend more time on them leading up to retirement. For me, that means getting a fix-me-up vintage motorcycle, learning at least one new programming language to be as proficient as my graduate students (unlikely), reducing the backlog on my Kindle and spending more time with family. Inherent in all these desires is the notion of not biting off more than one can chew. For if that is the case it is business as usual and hence no sense of retirement at all.

Rule 2: Consider the financial and scholarly implications

If I have to generalize, when looking at my colleagues in industry, government and academia in many parts of the world, money is, thankfully, often not the major factor in considering if and when to retire. They have frequently been working for a considerable period of time at institutions that have good retirement plans. Moreover, in keeping with the flexible notion of what "retirement" means there may be opportunities for paid part-time work not considered previously, such as being a consultant.

If you are one of the few people who are young and reading this, let me state what you have heard many times I am sure. Plan your financial future now. The wonders of compound interest should not be ignored. Financial security will definitely be part of your retirement calculus and that day comes sooner than you think.

Without money as a major concern *pro bono* activities are more likely to arise as an unpaid consultant, scientific advisory board member and so on. These activities also allow one to keep up to date with the latest science if one chooses. Although that can be harder if one does not have access to the closed access literature offered by universities etc. – a sad statement. Maintaining an adjunct affiliation with an institution that has a closed access journal subscription can address this, but there are often strings attached like requirements to teach.

Rule 3: Consider realistically what can yet be accomplished professionally

This could apply up until you retire but based on Rule 1 could continue well into the "retirement" years. If you are obsessed with accomplishing more, then probably best not to retire at all. There is also the question of, accomplish for whom? If it is for yourself that might lead to a different path than if it is in service to others.

For me, I think about it in terms of whether I have achieved my career goals.  In thinking deeply, I realize I never really had specific goals at the outset. Instead, they appeared as I lurched forward from researcher and added mentor, administrator and institution builder. For me it's about contentment - I feel content, that is what matters to me. What remains is to see things through as covered by other rules and to make my list of things yet to do. See Rule 8.

Rule 4: Consider your loved ones

I have told this story to students many times as part of a lesson in life-work balance. When my son was little he would sit on my lap and hold my head in his little hands and make eye contact. Some twelve years later my daughter did exactly the same thing. My immediate reaction was to put it down to their shared gene pool. Then one day I had a horrible realization - it was nurture, or lack thereof. Both my children, at many different times, had come up to me and said, "Dad, help me with…" and I, without my eyes leaving the computer screen, would say, "ah ha, ah ha."" For years, after that realization and now that they are adults, I have said to them, "Did you feel neglected?" They always respond, "Nah, or course not, Dad." I am not so sure, and that is something I want to address in the coming years. That I did not practice what I preach to my students is yet another matter.

Beyond that, and in deference to my wife of thirty-five years, is to redress the sacrifices she has made. The partner of a computational biologist, or any scientist for that matter, is one of sacrifice, both professionally and emotionally. One's partner has not only to love you, but to love your work as you do. Significant redress is long overdue.

Rule 5: Consider your colleagues and take steps before you retire

You are, of course, not obligated to continue working. However, there are many professional considerations that may affect the calculus: for example, if you do not retire, you may be preventing someone else from getting your tenured appointment; if you do retire, you may feel as if you are leaving students in the lurch, while leaving the courses that you teach and various administrative duties to fall to some poor soul and so on. Some of what you impart to colleagues is unique and will be lost, but that must be balanced against your own wishes and needs.

Try and mitigate what will be lost by nurturing others to take over those responsibilities ahead of time. Of course, this does not just apply at retirement, but also when you change jobs or responsibilities. The most important management and leadership skill I have learnt over the years is to plan for your successor the day you start a new position. This makes separation smoother.

Rule 6: Consider what your field will gain or lose and then ignore it

I am not attempting to speak for anyone but myself, but I think that it is far too presumptuous to think I am contributing anything unique. Over my forty-five-year career, science in general, and computational biology in particular, has become a team sport. The complexity of the work demands no less. We are each part of the solution, but not uniquely so. From the beginning, Isaac Newton's about "standing on the shoulders of giants" rings true. As time passes, individual contributions fade into the fabric that is science past and only in rare cases will individuals stand out in the eyes of future generations. I imagine that there are not a great many papers from the seventies or scientists in your field that you can cite or name from that era. Be satisfied to be part of those broad shoulders upon which future scientists will stand.

Rule 7: Consider your health and wellbeing

This is a no-brainer, ill-health can strike at any time. Less likely if you have maintained a regime of exercise, good diet, enough sleep, etc. over your lifetime. One question to ask is, if something happens health-wise and I am still working, will I have lost the opportunity to do other things that

I will now not be able to do. Stated another way, will my final words be, "I wish I had written just one more paper," or something else. Planning around that scenario would seem wise.

Rule 8: Consider opportunities not available otherwise and be prepared for the unexpected

As scientists we are inquisitive and life-long learners, yet we never have enough time for all the avenues, professional or otherwise, we wish to explore. For me, those things include one or two reviews that I think should be written, an autobiographical perspective on biomedical data (seriously) and a small number of research threads still to be pursued. All can be done at a different pace and likely with a different perspective than when heavily involved in running a research laboratory and institute. What's more they can be done easily. What may plague one during a career as a computational biologist – the ability to work at home as easily as working in the lab – now can become an asset. You can easily keep your hand in, but at your own pace and doing those tasks that one enjoys without the need to write grants or undertake a myriad of administrative tasks.

Beyond that, those retired or at least slowed down have told me that completely unexpected pursuits arise, either professionally or otherwise. I am excited to see what interesting opportunities arise and am ready to embrace the unexpected.

Rule 9: Consider the possibilities of giving back in new ways

This could take many forms and be rewarding. Examples that come to mind are various types of voluntary work, either working for or starting a foundation and mentoring at various levels. There is always a need for free expertise in science, technology, engineering and medicine (STEM). In fact, refer to Rules 1 and 2: consider how giving back could factor into your day-to-day retirement lifestyle.

Rule 10: Put your scholarly legacy in order

Part of your legacy is what you leave in the scholarly literature (noting Rule 6). Much of that is already in order thanks to our journals and practice of publishing. Yet there is more to one's scholarship, some of which might prove to be more useful than the papers themselves. There are data, software, laboratory protocols, course materials and so on. Are they available and in the best shape for reuse or for passing knowledge forward? At the point of retirement presumably one is less concerned about getting credit for hot new science and more about ensuring that knowledge is not lost. It is the liberating time in one's career to do what is right, not do what it takes to get promoted.

There you have it. Most of you probably did not get this far as you are not even close to reaching retirement and it seems ridiculous to contemplate. Trust me, when you do get to read to the end, you too will be well aware retirement comes quick enough. Whatever your later years bring, I hope they are as rewarding as I plan to make mine - someday.

If you did get to the end, perhaps you are retired and had the time and inclination to read these rules. Hopefully, you also have the time to comment on them from your own perspective, for as I said at the outset, first, these rules derive from someone who thinks about retirement, but who has not yet retired and second, everyone will have a unique perspective on the subject.

Acknowledgements:
Thanks to Fran Lewitter and David Searls for their perspectives on being "retired" and to Ira Klein and Teri Klein who were kind enough to weigh in with corrections and their own unique perspectives.